\documentclass[conference]{IEEEtran}
\usepackage{graphicx}
\usepackage{hyperref}
\usepackage{color,amsmath}
\usepackage{psfrag}
\usepackage{cite} 
\usepackage{amssymb, amsmath, amsfonts}
\usepackage{graphics}
\usepackage{fancyhdr}
\usepackage{eucal}
\usepackage[english]{babel}
\usepackage{subfigure}

\newif\ifnotes\notestrue

\def\hgr#1{}

\begin{document}
\title{Single--sided Real--time PESQ Score Estimation%
\thanks{Partially supported by the EuroNF ``QUAXIME'' project.}
\thanks{This work was partially supported by Celtic Easy Wireless 2 project, funded in Finland by Tekes (Finnish Funding Agency for Technology and Innovation) and VTT}
} 

%
%
\author{Sebasti\'an Basterrech, Gerardo Rubino \\INRIA/Rennes -- Bretagne Atlantique, \\ Rennes, France \and  Mart\'\i n Varela \\ Converging Networks Laboratory\\VTT Technical Research Centre of Finland \\ Oulu, Finland}
\date{\empty}
\maketitle
\thispagestyle{empty}

\begin{abstract}
For several years now, the ITU-T's Perceptual Evaluation of Speech
Quality (PESQ~\cite{itu:p.862}) has been the reference for objective
speech quality assessment. It is widely deployed in commercial QoE
measurement products, and it has been well studied in the literature.
While PESQ does provide reasonably good correlation with subjective
scores for VoIP applications, the algorithm itself is not usable in a
real--time context, since it requires a reference signal, which is
usually not available in normal conditions. In this paper we provide
an alternative technique for estimating PESQ scores in a single--sided
fashion, based on the PSQA technique~\cite{perfeval}.
\end{abstract}

\section{Introduction}
\label{sec:introduction}
%
%
For several years now, the ITU-T's Perceptual Evaluation of Speech
Quality (PESQ~\cite{itu:p.862}) has been the reference for objective
speech quality assessment. It is widely deployed in commercial QoE
measurement products, and it has been well studied in the literature.\\

In previous work~\cite{varela:mesaqin06}, we have studied the
performance of PESQ for VoIP over a wide range of network conditions,
and found that
\begin{enumerate}
\item the correlation with subjective scores was good, even for cases
  in which losses were relatively abundant and bursty (but still
  within reasonable limits, see ~\cite{pennock1,psytechnics:pesq} for
  some limitations of PESQ with respect to network impairments), and
\item PESQ scores were fairly consistent for all combinations of
  speech samples and loss patterns.\\
\end{enumerate}

These results lead to thinking that a good approximation of PESQ can
be achieved at the receiving terminal as long as network performance
can be measured and some application--level knowledge (such as the
codec in use, the presence or absence of loss concealment, etc.) is
available. In the past we have advocated the use of Pseudo--Subjective
Quality Assessment (PSQA~\cite{varela:thesis}) for VoIP QoE
estimations, which allows for very accurate estimations of MOS values
based on network and application parameters. In this paper we analyze
the applicability of the PSQA approach to the estimation of PESQ scores.\\

In principle, given that both PESQ and PSQA correlate very well with
subjective perception, it is expected that the approach presented
herein will lead to a hybrid approach offering the best of both
worlds. On the one hand, using PESQ as a target function eliminates
the costly part of PSQA, namely the need to perform a non--trivial
subjective LQ assessment campaign. On the other hand, it allows to
have these results in real--time, without the need for a reference
signal. This enables the use of these PESQ-like results in situations
in which some quick reaction is desirable, for instance in order to
improve the perceived quality by means of real--time controlling
actions on the communication system whose delivered QoE is
automatically assessed, one of the main goals behind our
research efforts.\\

The paper is organized as follows. In
  Section~\ref{sec:methodology} we describe the experiments realized
  and their motivation.  Section~\ref{sec:results} presents the
  results obtained. We conclude the paper in
  Section~\ref{sec:conclusions}.

\section{Methodology}
\label{sec:methodology}

\subsection{Motivation}
\label{sec:motivation}
Our previous work on PSQA is based on a rather simple concept, to wit: the
quality of a media stream (be it voice or video), as perceived by an
average user, and assuming no extraneous, non-measurable degradations
at the source (such as faulty equipment) is usually determined by a
number of factors that can be divided in two categories. These categories are
\begin{description}
\item[] \textbf{Application--related factors}, such as the encoding used, the
  type of error correction and concealment chosen, play-out buffer
  sizes, etc.
\item[] \textbf{Network--related factors}, such as the loss rate in the network,
  delay, jitter, etc.\\
\end{description}

These premises, coupled with the fact that PSQA provides very good
correlation with subjective scores, imply a certain independence of
the perceived quality from the actual media streamed (there are of
course some limitations to this claim, especially concerning video,
mostly related to scene types and amount of motion, but those can be
measured and hence considered as an application--level factor).\\

In turn, the previous observation leads to the prediction that for
VoIP, the scores given by reference--based tools such as PESQ should
be quite consistent for a given configuration of application and
network factors or parameters. This was the subject of our work
in~\cite{varela:mesaqin06}. The results from that study show that PESQ
scores taken for a single encoding and over consistent network conditions
are remarkably stable. So much so, that for a given configuration of
parameters (in the case of the previous study, a given codec, whether
PLC was in use and the loss rate and loss distribution in the network)
a fairly good prediction of the PESQ-LQ values could be given by
taking the median of a series of PESQ-LQ assessments taken over
similar configurations. For \emph{reasonable} network conditions
(i.e. conditions that do not degrade the VoIP stream's quality badly
enough to break PESQ's assessment), the median--based estimations are
very close to actual PESQ scores. \\

Using this approach in practice, however, has some
limitations. Firstly, it requires a rather large number of assessments
to be performed in order to acquire enough information to reliably
cover the parameter space. This, in itself is not a serious issue if
the parameters considered are not too numerous, but it is an area that
could be improved. The second issue is more important, since it may
actually limit the applicability of the approach. This issue is the
lack of generalization and hence the inability to extrapolate for
parameter values not present in the original measurements. While this
could be palliated by a brute--force approach (i.e. cover a larger
parameter space, in a more fine--grained fashion if needed), this is
not an elegant solution, and it basically doesn't solve the issue, but
only masks it.\\

PSQA, on the other hand, relies  upon the ability of the
Neural Network (NN) it uses as a learning tool in order to reduce
the number of samples required to reliably cover the whole parameter
space. This is important since PSQA is usually trained with subjective
scores, which are expensive and time--consuming to obtain. The NN's
ability to generalize, coupled with PESQ's regularity over similar
application and network configurations hint at the feasibility of
obtaining a flexible, cheap and accurate way of single--sidedly
estimating PESQ scores by using PSQA.\\

\subsection{Experimental Setup}
\label{sec:setup}

The experimental setup used for this study is very similar to
the one used for~\cite{varela:mesaqin06}. We used G.711 encoding, with
and without loss concealment, and considered the loss rate and
distribution in the network as our network parameters. While jitter is
a relevant parameter for LQ, it can be folded into the loss rate if no
particular attention is being payed to the dejittering buffer sizes
and algorithms. Hence, it is not considered explicitly in this study.\\

The network loss model used is a simplified Gilbert
model~\cite{gilbert1} in which the lossy--state loss probability is 1
(i.e. all packets are lost in the lossy state). This model has the
advantage of eliminating one free variable, and it provides a
reasonably good model of losses on the Internet.\\

The network impairments are thus represented not only by the packet
loss rate~(LR), but also by the dispersion of losses in the stream,
captured by the mean loss burst size~(MLBS)~\cite{perfeval}.  The
MLBS is the expected number of consecutive losses in a loss episode,
that is, the mean length of loss bursts in the flow, a real number
$\geq 1$.  We considered loss rates between 1\% and 30\%,
more specifically values 1\%, 2\%, 3\%,~..., 30\%,
and mean loss burst sizes of up to 6 consecutive packets (values
1, 1.25, 1.5, 1.75, 2,
2.5, 3, 3.5, 4, 5, 6).  Given that standard--length (approximately 10s)
samples were used, it was not possible to have all possible
combinations of LR and MLBS, since some of them are not really
feasible within the $\sim 400$ packets that each speech sample uses
when transmitted. Thus, only valid combinations were considered, and
for those, each loss trace created was verified to ensure that it had
the desired characteristics.\\

It should also be noted that PESQ is not expected to behave
  correctly with respect to subjective scores when the network
  impairments are too high. In any case, since the goal of the study
  was to mimic PESQ's performance, we anyway considered very impaired
  networks.\\

  For each combination of values of the two loss-related parameters LR
  and MLBS, 10 different traces (all statistically similar) and 20
  standard speech samples (50\% male and 50\% female) were
  used\footnote{For some configurations in the higher--end of the
    impairment values we actually used more samples, in order to
    mitigate the variability of PESQ's results.}.  The number of
  samples generated and then evaluated with PESQ was slightly above
  128500.  For each combination of LR, MLBS and packet loss
  concealment (PLC, either active, coded PLC = 1, or not, coded PLC =
  0) several sequences were analyzed (around 200 of them, except in
  some cases with high loss rates, where more samples were generated
  and used). In other words, we sent each one of the error-free voice
  sequences through a simulated/emulated network varying the three
  considered variables, and we used PESQ to evaluate the resulting
  quality.  Since with every considered triple of values for LR, MLBS,
  PLC (we call a \textit{configuration} such a
  triple~\cite{varela:thesis}) we had many different associated PESQ
  values, we generated a second smaller table having around 600
  entries, each corresponding to a different configuration of our
  platform. Again, in this table, each considered configuration (a
  loss rate, a value for the mean loss burst size, and the indicator
  of packet concealment active or inactive), there is one row in this new table. \\
  
  For each of the entries (configurations) of the compact table, we
  evaluated statistical descriptors of the set of PESQ values
  associated with, such as the empirical mean, median, variance,
  etc. As in \cite{varela:mesaqin06}, the median was a good
  approximation of PESQ scores. We therefore used it to train a Neural
  Network using the AMORE package for the R statistical analysis
  language. That is, we built a function $f$ mapping each possible
  configuration into a quality value in the interval $[1,5]$
  (actually, given that the target function is PESQ, the interval will
  be $[1,4.5]$), that approximates the median of the values obtained
  using PESQ. Function $f$ is defined in the space
  $[1,30]\times[1,6]\times\{0,1\}$, corresponding to LR in \%, MLBS
  and PLC. This function $f$ is our approximation tool for PESQ, whose
  behavior is analyzed in next Section.\\

\section{Results}
\label{sec:results}

The learning phase consisted of using a standard Neural Network (NN)
for learning the mapping from configurations to (median) PESQ values.
This was also partly done in the context of a larger study
  comparing the performance of the AMORE--based NNs against the Random
  Neural Networks (RNN) we have used previously. This comparison work
  is still ongoing at the time of
  writing.
  Some preliminary results were published in~\cite{psqa:icann06}; for the tool
  itself and its use in the PSQA approach, see~\cite{grChapter05}.
  Any of the numerous good references on Neural Network methodology
can provide background material to the reader if this is necessary;
for a classic one, see~\cite{bishop06}. \\

For training the NN, we randomly (and uniformly) separated the data in
the small compacted table into two parts, corresponding to a
80\%--20\% decomposition for training and validation respectively.
Since we have a binary variable PLC in the configurations, we actually
built 2 NN, that is, two functions, $f_{0}$ corresponding to the case
PLC$ = 0$, and $f_{1}$ for the case of PLC$ = 1$. This proved to
  be a better solution in this case than having a single NN with the
  PLC considered as a third input.\\

We used the usual 3-layer feed-forward perceptron structure
with two inputs (LR and MLBS)
and one output (estimated, or predicted PESQ value).
For the hidden layer, we varied the number of
neurons starting from~1, in order to select the best architecture for
our neural networks.
We finally chose an architecture with~30 hidden neurons for both
$f_{0}$ and $f_{1}$. As stated before, the whole data set
for learning (coming from the small table) has
around 600 entries, half corresponding to the case PLC$=0$
and half for the case PLC$=1$.\\

Let us denote
by ${\cal TS},i$ the set of configurations corresponding to
the 80\% used for training the $f_{i}$ network,
the \textit{Training Set} for the case PLC = $i$, with
cardinality $K_{{\cal TS},i}$, and by
${\cal VS},i$ the similar set of configurations corresponding to
the 20\% used for validation (the \textit{Validation Set} when PLC = $i$),
with cardinality $K_{{\cal VS},i}$.
The \textit{Training Error} when PLC = $i$, $i = 0$ or $1$, is then
	$$ (K_{{\cal TS},i})^{-1} \!\!\!\!\!\!\!\!\!\!\!\!
		\sum_{\text{all config.} \gamma \in {\cal TS},i}
		\!\!\!\!\!  \big[ f_{i}(\gamma) - \text{MedianOfPESQ}(\gamma) \big]^2,
	$$

and the \textit{Validation Error} is
	$$ (K_{{\cal TV},i})^{-1}  \!\!\!\!\!\!\!\!\!\!\!\!
		\sum_{\text{all config.} \gamma \in {\cal VS},i}
	\!\!\!\!\! 	\big[ f_{i}(\gamma) - \text{MedianOfPESQ}(\gamma) \big]^2.
	$$
In both expressions, we call configuration (denoted by $\gamma$) just the pair
(LR,MLBS), since we separated the data into two sets thus eliminating the need
for a third variable PLC. For each such $\gamma$, 
MedianOfPESQ$(\gamma)$ is the value obtained from the analysis of
the original table having fixed PLC to 0 or to 1, according to the case
we are analyzing, for instance, the number defined by
	$$\text{argmin}_{x} K^{-1} \!\!\!\!\!
		\sum_{\text{all config. } \gamma}
		\!\!\!\!\! \big| \text{PESQ}(\gamma) - x \big|
	$$
if $K$ is the size of the small table (around 600 in our experiments). 	
Table~\ref{t:NN} provides some data for
this step of the analysis. Given the fact that we are using PESQ values in the
range [1,5], the reached error levels are indeed extremely small.

\begin{table}[htdp]
\caption{Performances of the learning phase,
for the two selected Neural Networks $f_{0}$ and $f_{1}$}
\begin{center}
\begin{tabular}{c||c|c}\hline
neural & training & validation \\
network & error & error \\ \hline
$f_{0}$ & $0.064$ & $0.069$ \\
$f_{1}$ & $0.040$ & $0.042$ \\
\hline
\end{tabular}
\end{center}
\label{t:NN}
\end{table}

\begin{figure*}
\begin{center}
\fbox{
\includegraphics[angle=-90,scale=0.5]{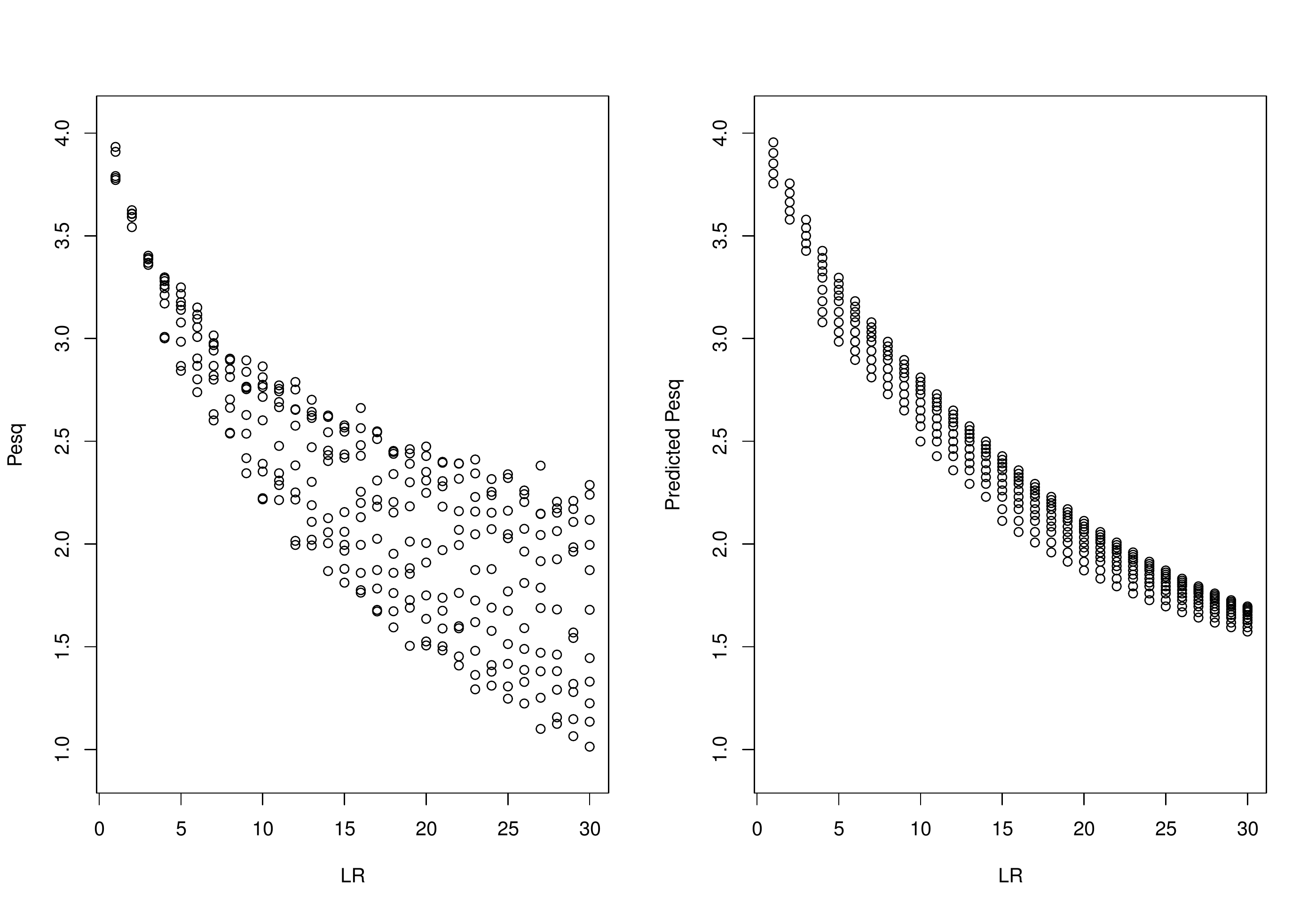} 
}
\caption{\label{PesqVsLrnoPlc}Case of PLC$=0$. PESQ and its predictor $f_{0}$, as an explicit function of LR.
Each spot corresponds to a specific configuration in the small table. Different spots
for a same LR correspond to different values for MLBS.}
\end{center}
\end{figure*}

Figure~\ref{PesqVsLrnoPlc} shows, on the left, PESQ values and on
  the right, the predictions provided by the Neural Network,
  everything for PLC~=~0 (no Packet Loss Concealment). In the $x$-axis
  we put LR values. Each point in the graphs corresponds to a
  configuration (LR,MLBS,0). Different points on the same vertical
  line, that is, with same LR, correspond to configuration with same
  LR but varying MLBS. It is interesting to see that the PESQ plot
  shows a significant amount of dispersion compared to the estimation
  when the loss rate goes over 10 to 15\%.  This is due to the NN
  being trained with median values, which significantly suppress the
  impact of outlier values in the data set. It is also known that PESQ
  tends to behave in a more variable way when the network impairments
  become large, and this behavior is exacerbated in this case by the
  lack of PLC on the decoder end.\\


\begin{figure*}
\begin{center}
\fbox{
\includegraphics[angle=-90,scale=0.5]{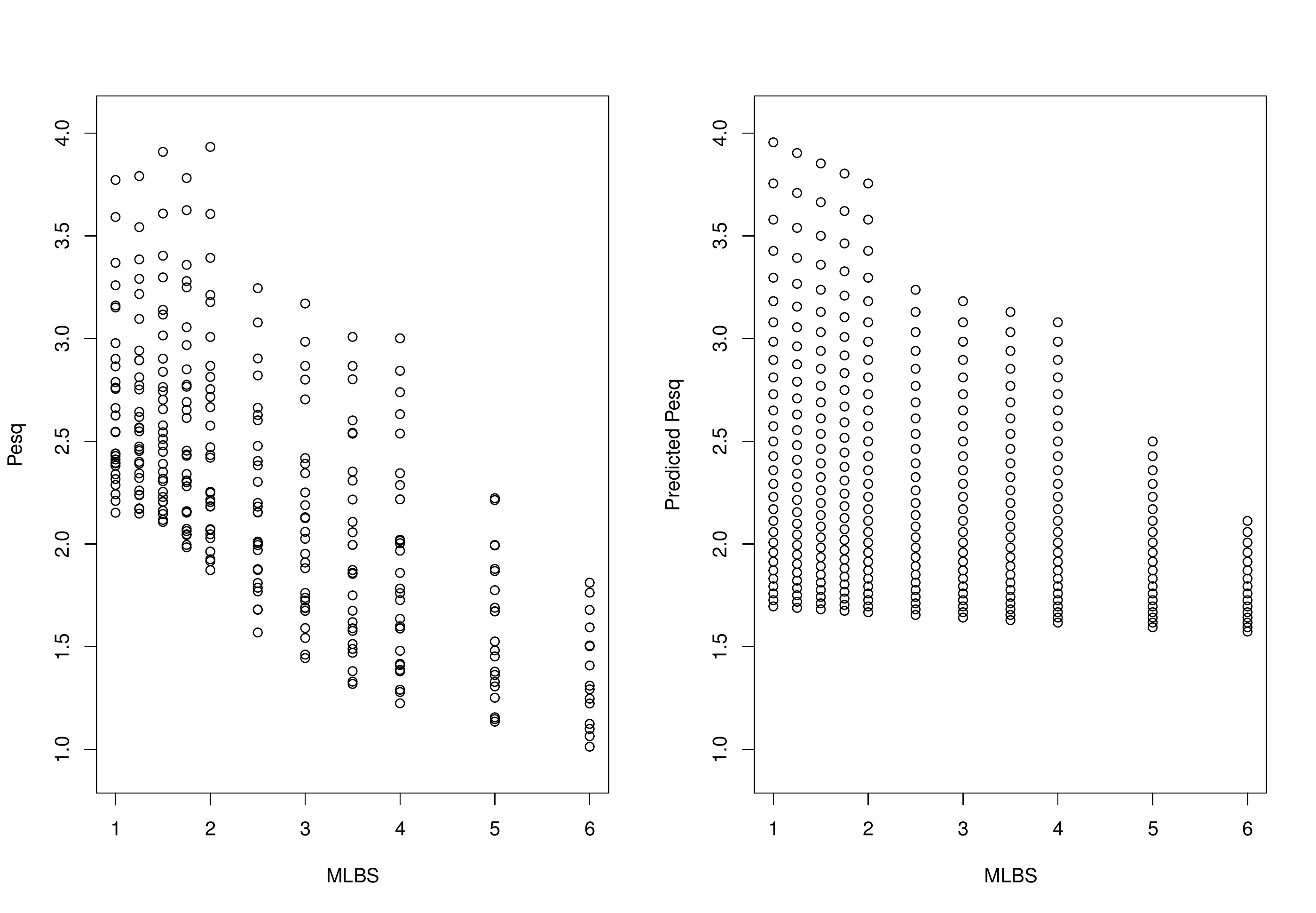} 
}
\caption{\label{PesqVsMLBSnoPlc}Case of PLC$=0$. PESQ and its predictor $f_{0}$, as an explicit function of MLBS.
Each spot corresponds to a specific configuration in the small table. Different spots
for a same MLBS correspond to different values for LR.}
\end{center}
\end{figure*}

Figure~\ref{PesqVsMLBSnoPlc} provides an analogous view, plotting
  PESQ and its estimation as a function of MLBS. It can be noticed in
  this plot that the estimated values are not as expected for burst
  losses higher than two or three packets, in which case the
  estimations are overly optimistic with respect to actual PESQ
  values.  We do not, at the time, have a definitive explanation for
  this phenomenon. However, given the good correlation for PSQA and
  subjective scores obtained in previous study, we suspect that the
  variability of PESQ results with respect to MLBS might have
  precluded the NN from capturing the correct behavior.\\

\begin{figure*}
\begin{center}
\fbox{
\includegraphics[angle=-90,scale=0.5]{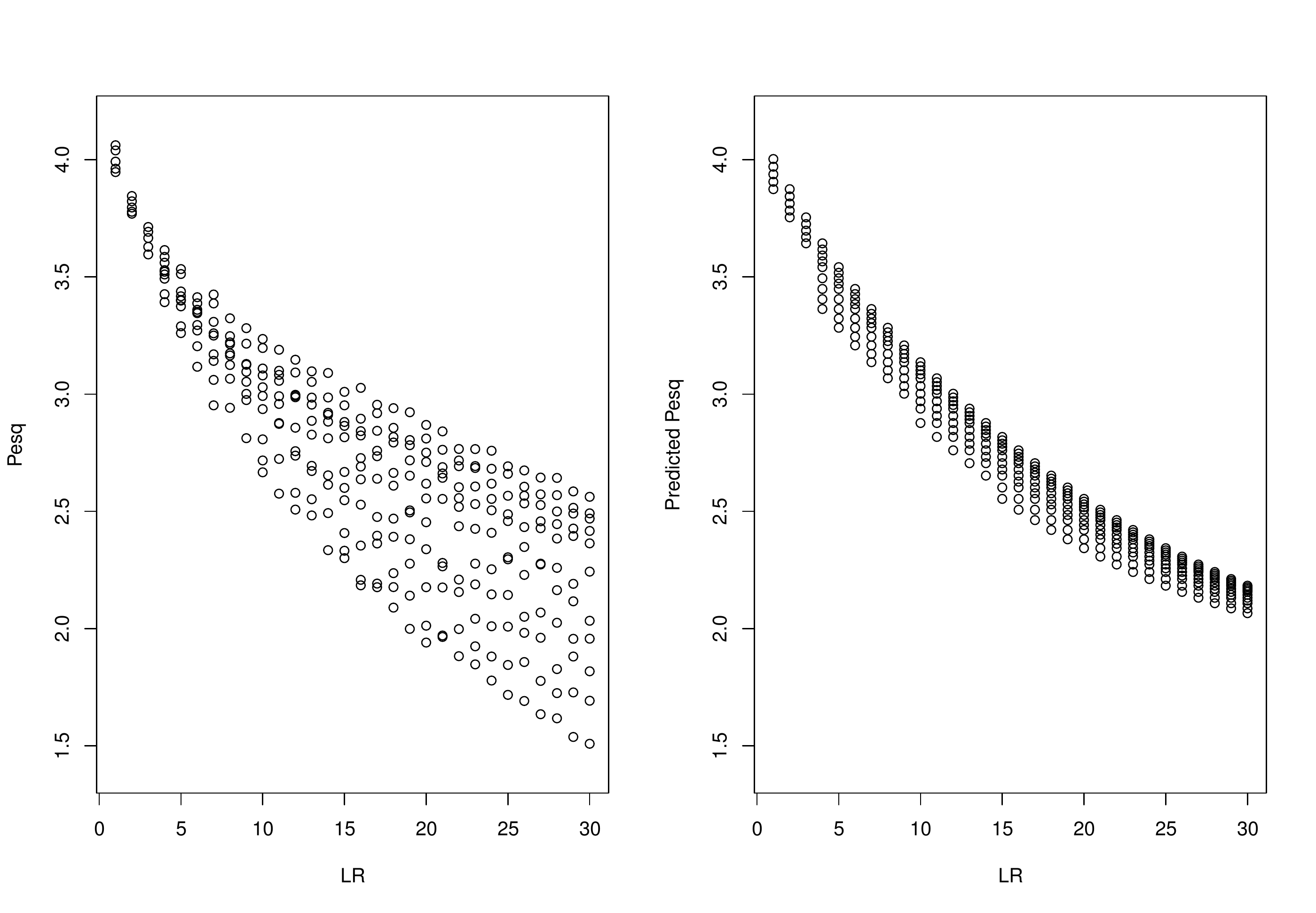} 
}
\caption{\label{PesqVsLrPlc}Case of PLC$=1$. PESQ and its predictor $f_{1}$, as an explicit function of LR.
Each spot corresponds to a specific configuration in the small table. Different spots
for a same LR correspond to different values for MLBS.}
\end{center}
\end{figure*} 
\begin{figure*}
\begin{center}
\fbox{
\includegraphics[angle=-90,scale=0.5]{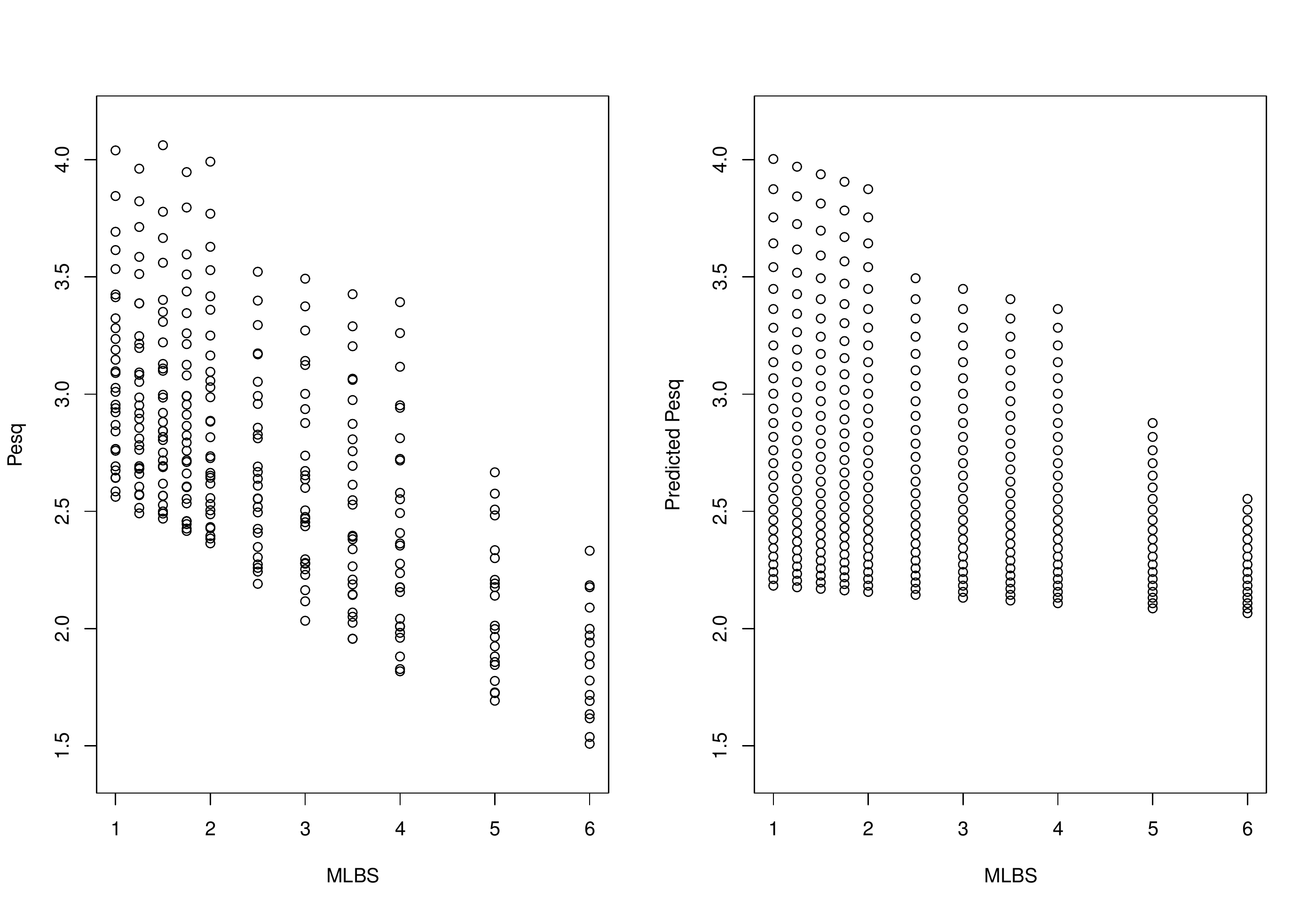} 
}
\caption{\label{PesqVsMLBSPlc}Case of PLC$=1$. PESQ and its predictor $f_{1}$, as an explicit function of MLBS.
Each spot corresponds to a specific configuration in the small table. Different spots
for a same MLBS correspond to different values for LR.}
\end{center}
\end{figure*}


Figures~\ref{PesqVsLrPlc} and~\ref{PesqVsMLBSPlc} illustrate the
  case of PLC~=~1. As expected, the overall values in this case are
  higher (by about 0.5 MOS points) than in the non-PLC
  scenario. Otherwise, the overall behavior of PESQ and the NN--based
  estimations are comparable to the non--PLC case, but with smaller
  errors.\\

Consider again the original set of data (the large table),
over~$10^5$ voice
  samples, with the corresponding values of loss rate, mean loss
  burst size and PLC, together with the quality assessment made by PESQ. If we
  use our functions $f_{i}$ for approximating the PESQ scores for all
  of the data points, what would be the mean error?
 Observe that this is quite close of a field application of our approach,
 even if this table of values is the original data with which the training
 data sets were built.
 Denote
 
\begin{itemize}

\item by~$s$ a generic entry in the original table (a sample); there are
more than $10^5$ such samples;

\item by PLC$(s)$ the value of PLC in sample~$s$;

\item by $f_{\text{PLC}(s)}(s)$ the value
  predicted by the right NN when the configuration is the one in sample~$s$;
  
\item finally, let PESQ$(s)$ be the PESQ assessment of sample~$s$.

\end{itemize}

  Table~\ref{t:global} shows the Mean Square Error (MSE$_i$), its square
  root and the Mean Absolute Error (MAE$_i$), corresponding to
  function $f_{i}$, defined as follows:
	$$\text{MSE}_{i} = \frac{1}{N_{i}} \sum_{ s: \text{PLC}(s)=i }
		\big[ f_{\text{PLC}(s)}(s) - \text{PESQ}(s) \big]^{2},
	$$
	$$\text{MAE}_{i} = \frac{1}{N_{i}} \sum_{ s: \text{PLC}(s)=i }
		\big| f_{\text{PLC}(s)}(s) - \text{PESQ}(s) |.
	$$\\

\begin{table}[htdp]
\caption{Performances of the two selected Neural Networks $f_{0}$ and $f_{1}$}
\begin{center}
\begin{tabular}{c||c|c|c}\hline
neural & \raisebox{-1ex}{MSE} & \raisebox{-1ex}{$\sqrt{\text{MSE}}$} & \raisebox{-1ex}{MAE} \\
network & ~ & ~ & ~ \\ \hline
$f_{0}$ & $0.236 $ & $0.486$ & $0.412$ \\
$f_{1}$ & $0.076$ & $0.276$ & $0.221$ \\
\hline
\end{tabular}
\end{center}
\label{t:global}
\end{table}

\begin{figure*}
\begin{center}
\fbox{
\includegraphics[angle=-90,scale=0.6]{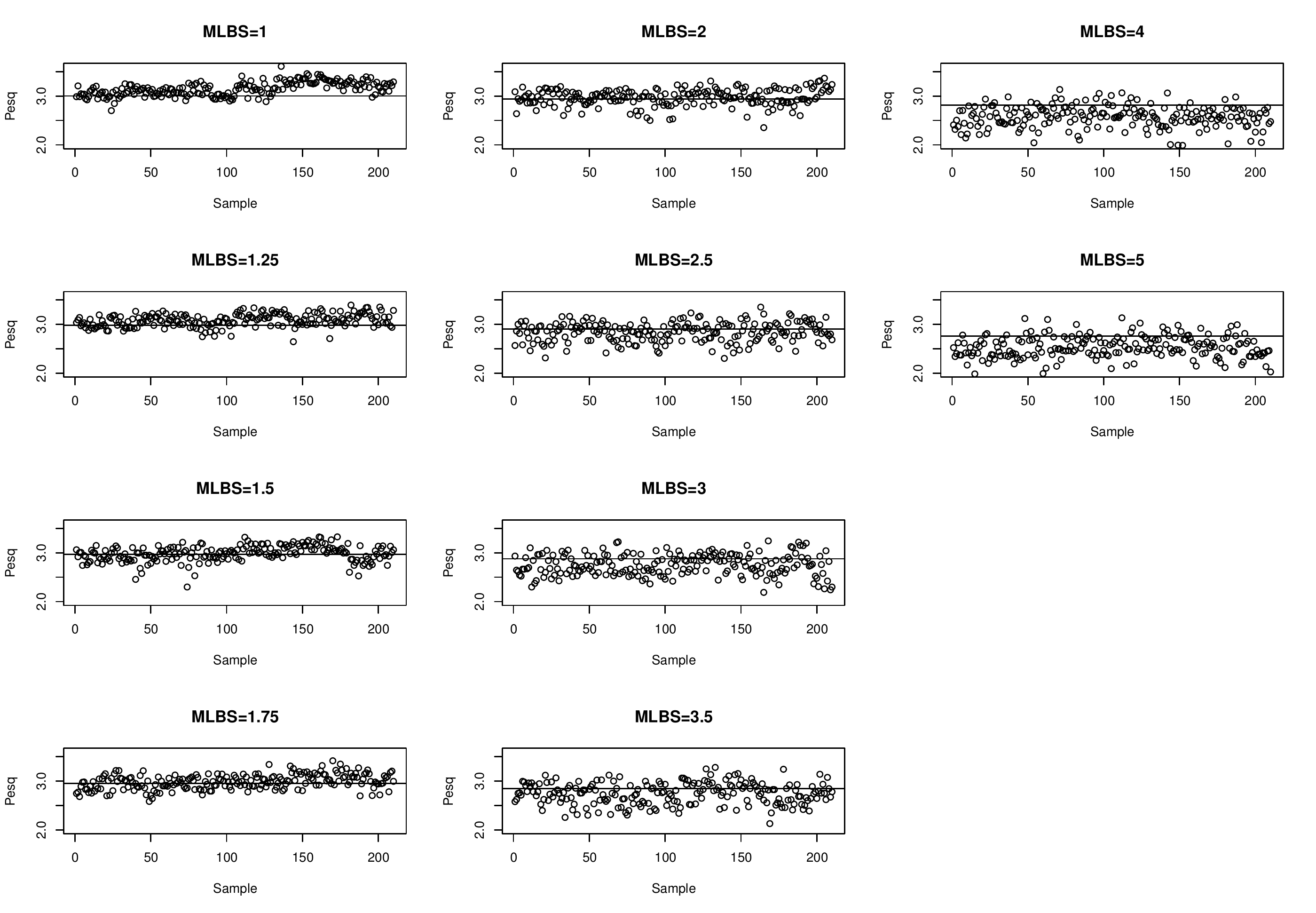} 
}
\caption{\label{fig:approximations}PESQ values for all points in the data set, where LR=12. Each area represents a separate MLBS value, and the horizontal lines represent the NN's estimation of the PESQ score. PLC=1.
}
\end{center}
\end{figure*}

This implies that the NN--based estimations are on average, at about
0.41 points from actual PESQ scores for samples in which PLC was not
used, and at about 0.22 points for samples in which it was
enabled. Given the average listener's appreciation in terms of the MOS
scale, it seems that the estimations are indeed very close to the
actual values. This closeness can be seen in Figure~\ref{fig:approximations}, which shows, for a loss rate of 12\%
all the PESQ scores in the complete data set, separated by MLBS value, and the NN's estimation of them.\\

\section{Conclusions}
\label{sec:conclusions}

In this paper we have presented a simple, efficient way of
  providing single--sided, reference--free estimations of PESQ scores
  for VoIP samples or ongoing streams. The method used was PSQA
  (Pseudo--Subjective Quality Assessment), but using PESQ as a target
  function instead of subjective scores, as was done previously.\\

  While this will evidently not increase the correlation of PSQA with
  respect to subjective scores, it provides a very cheap and efficient
  way of having a single--sided quality assessment tool. Moreover, the
  evaluation by NNs is very computationally efficient, which allows
  this mechanism to be used in real--time, for control purposes, for
  example, even in resource--constrained devices such as mobile phones
  or Internet tablets (unlike, say, the ITU's P.563~\cite{p.56304:_singl_ended_method_objec_speec} single sided metric, which is very resource--intensive).\\

  The reliability of the results obtained is slightly variable with
  network conditions, as depicted in Figures~\ref{PesqVsLrnoPlc}
  through~\ref{PesqVsMLBSPlc}. However, it should be noted that
  firstly, for \emph{normal} operating conditions, in which network
  impairments are not too high, the estimations are remarkably close
  to actual PESQ scores. Secondly, since PESQ itself shows reliability
  issues in cases where the network is severely impaired, a different
  approach should be tried in these scenarios, as needed.\\

  In future work on this subject, we plan on determining the
  impairment bopundaries in which using this sort of approach works
  well in practice, and using it to implement some sort of QoE control
  mechanism (either application or network--based). It would be also
  interesting to use different kinds of neural networks (for example
  in a recurrent architecture, instead of feed--forward) and also to
  re--use the data obtained in this work to train a Random Neural
  Network (RNN, cf~\cite{gelenbe:random}), which we have previously
  used with success for PSQA applications.

\bibliographystyle{unsrt} 
\bibliography{rapport}
\end{document}